\documentclass[12pt,preprint]{aastex}

\shorttitle{Plasma Brightenings in a Failed Filament Eruption}
\shortauthors{Li et al.}

\begin{document}

\title{Plasma Brightenings in a Failed Solar Filament Eruption}

\author{Y. Li$^{1,2,3}$, M. D. Ding$^{1,3}$}
\affil{$^1$School of Astronomy and Space Science, Nanjing University, Nanjing 210023, China; yingli@nju.edu.cn}
\affil{$^2$Department of Physics, Montana State University, Bozeman, MT 59717, USA}
\affil{$^3$Key Laboratory for Modern Astronomy and Astrophysics (Nanjing University), Ministry of Education, Nanjing 210023, China}

\begin{abstract}
Failed filament eruptions are solar eruptions that are not associated with coronal mass ejections. In a failed filament eruption, the filament materials usually show some ascending and falling motions as well as generate bright EUV emissions. Here we report a failed filament eruption that occurred in a quiet-Sun region observed by the Atmospheric Imaging Assembly on board the {\em Solar Dynamics Observatory}. In this event, the filament spreads out but gets confined by the surrounding magnetic field. When interacting with the ambient magnetic field, the filament material brightens up and flows along the magnetic field lines through the corona to the chromosphere. We find that some materials slide down along the lifting magnetic structure containing the filament and impact the chromosphere to cause two ribbon-like brightenings in a wide temperature range through kinetic energy dissipation. There is evidence suggesting that magnetic reconnection occurs between the filament magnetic structure and the surrounding magnetic fields where filament plasma is heated to coronal temperatures. In addition, thread-like brightenings show up on top of the erupting magnetic fields at low temperatures, which might be produced by an energy imbalance from a fast drop of radiative cooling due to plasma rarefaction. Thus, this single event of failed filament eruption shows existence of a variety of plasma brightenings that may be caused by completely different heating mechanisms.
\end{abstract}


 \section{Introduction}

Eruptions of solar filaments or prominences \citep{tand74} are one of the drastic activities in the solar atmosphere \citep{pare14,vial15,mcca15}. They involve energy release, plasma heating, as well as ascending and/or falling motions of plasma, which are coupled with the ambient magnetic field in both active regions and quiet-Sun regions. There are three kinds of filament eruptions \citep{gilb07}, referred to as full, partial, and confined (or failed; \citealt{jihs03}) ones. The former two are usually associated with a coronal mass ejection; while in the latter two, one might observe the filament material falling back to the Sun.

When the erupted filament material falls back to the Sun, it could cause bright impacts on the solar surface by kinetic energy dissipation. \cite{real13} reported the hot impacts after the 2011 June 7 eruption that was associated with an M2.5 flare. They found through a numerical simulation that the strong impact brightening comes from the original fragment material when hitting the solar surface. In their simulation, they did not consider the role of magnetic field, since the magnetic field in most of the impact regions (quiet-Sun) is rather weak, and most of the fragments are basically in free fall. \cite{petr16} studied the same event but focused on the impacts close to active regions with relatively strong magnetic fields. They observed that the magnetic channel brightens before the fragment impacts the solar surface. By performing a magnetohydrodynamic simulation with the magnetic field included, they found that the filament material is channelled by the magnetic field and that the brightening ahead of the pre-impact fragments mainly comes from a shocked region. \cite{inne16} also investigated the same event and reported that many of the falling fragments are dissipated above the chromosphere and cause brightenings in the corona, either due to ionization and trapping in magnetic fields or because of encountering a splash from earlier impacts.

The kinetic motion and emission property of the filament material could be influenced by the surrounding magnetic field. Firstly, the magnetic field could spread out, deform, and fragment the filament material \citep{inne12,petr16}. The filament material could also be slowed down by magnetic drag forces \citep{dole14} and even prevented from falling deep into the chromosphere \citep{inne16}. In some cases, for example, the magnetic pressure balances the ram pressure of filament material, or the material is ionized and so trapped in the magnetic field, the material would be channelled along the magnetic field lines \citep{petr16}. During the channelling, the material can generate bright EUV emission by kinetic energy dissipation \citep{inne16} or impact ionization \citep{down13}. In other words, the bright EUV emission of the material could simply trace out the magnetic field lines \citep{down13,inne16,petr16}. Note that, even in the quiet-Sun region, the confinement and impact of the falling material could produce loop-like brightenings, implying that the quiet-Sun corona is still structured by the magnetic field \citep{inne16}. In addition, bright EUV emissions could show up as a result of magnetic reconnection between the magnetic field in the filament and that in the ambient corona as manifested by a redirection of the filament material \citep{vand14,lilp16}.

In this paper, we report a failed filament eruption that occurred in a quiet-Sun region (i.e., confined eruption of a quiescent filament; \citealt{prie89}). In this event, the erupting filament is confined by the surrounding magnetic field. When interacting with the ambient magnetic field, the filament material brightens up and flows along the magnetic field lines through the corona to the chromosphere. We find two ribbon-like brightenings near the solar surface, plasma heating in some helical structures, and also thread-like brightenings on top of the erupting magnetic fields. These brightenings appearing in different magnetic structures in this event are possibly related to different kinds of mechanisms that produce local heating effect.

\section{Observations} 

\begin{figure*}
\centering
\includegraphics[width=13cm]{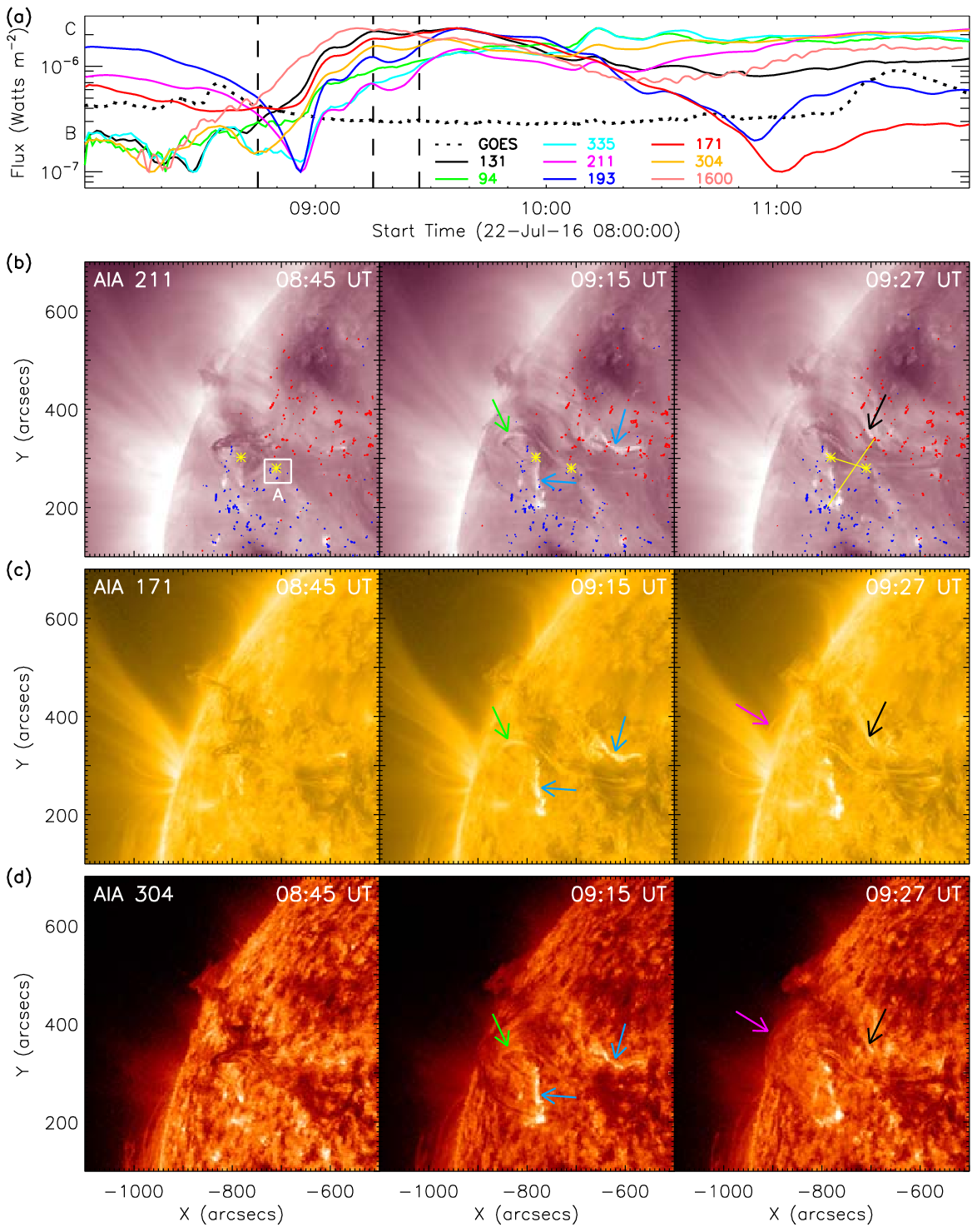}
\caption{{\small Overview of the failed filament eruption. (a) Light curves in the {\em GOES} 1--8 \AA~for the whole Sun and in multiple AIA passbands integrated over the region as shown in panels (b)--(d). (b)--(d) AIA 211 \AA, 171 \AA, and 304 \AA~images at three times indicated by the vertical dashed lines in panel (a). The blue arrows indicate the ribbon-like brightenings. The green, magenta, and black arrows mark some bright threads. In panel (b), the red and blue contours represent the magnetic polarities at $+$100 and $-$100 G, respectively. A pair of yellow asterisks mark the approximate position of the sheared filament arcade footpoints. The white box A is the same as the one in Figure \ref{fig-rec}(c). The two yellow lines in the image at 09:27 UT connect the possible conjugate footpoints of the sheared filament arcade and some less-sheared coronal field lines.}}
\label{fig-obs}
\end{figure*}

The event presented here is a failed filament eruption on 2016 July 22 in a quiet-Sun region near the northeast limb. It was observed by the Atmospheric Imaging Assembly (AIA; \citealt{leme12}) with a high spatial resolution ($0.\!\!^{\prime\prime}6$ pixel$^{-1}$) and time cadence (12 or 24 s) in the EUV\footnote{The EUV passbands include 131 \AA~($\sim$10 MK), 94 \AA~($\sim$6.3 MK), 335 \AA~($\sim$2.5 MK), 211 \AA~($\sim$2.0 MK), 193 \AA~($\sim$1.6 MK), and 171 \AA~($\sim$0.6 MK) for coronal plasmas, and 304 \AA~($\sim$0.05 MK) for chromospheric plasmas.} and UV\footnote{The UV 1600 \AA~passband contains the chromospheric continuum plus the transition region C {\sc iv} line.} passbands and by the Helioseismic and Magnetic Imager (HMI; \citealt{scho12}) on board the {\em Solar Dynamics Observatory} ({\em SDO}). From the multi-wavelength (or multi-temperature) AIA images (Animation 1 accompanying Figure \ref{fig-obs}), we can see that the filament spreads out but gets confined by the surrounding magnetic field. When interacting with the ambient magnetic fields, the dark material gets brightened in the form of threads (see the green and magenta arrows in Figure \ref{fig-obs}) as well as slides down along the lifting coronal field lines. As the material hits the chromosphere, it creates two ribbon-like brightenings (marked by the blue arrows in Figure \ref{fig-obs}). From the line of sight HMI magnetogram (see the contours at $\pm$100 G in Figure \ref{fig-obs}(b) and also Animation 1), it is seen that this event takes place in a bipolar region with weak magnetic fields and the polarity inversion line generally extends from northeast to southwest. Note that there exist some quite weak positive magnetic polarities within the main negative polarity area, and vice versa.

We notice that the erupting filament material likely flows along the magnetic field lines, displaying a sheared arcade structure containing the filament (referred to as sheared filament arcade hereafter) in an early time and a series of less-sheared thread structures at a later time (see the two yellow asterisks in Figure \ref{fig-obs}(b) marking the approximate position of the sheared filament arcade footpoints and the two yellow lines implying a change of shear in the magnetic field lines, i.e., the angle between the connecting line of conjugate footpoints and the polarity inversion line along the northeast--southwest direction becoming larger over time)\footnote{There should be a line of sight effect in the shear, but combining the HMI magnetogram, we could say that a sheared arcade structure and some less-sheared field lines show up in this region.}. Note that some background coronal arcades could also be seen in the AIA 211 \AA~and 193 \AA~images. As the sheared filament arcade expands and interacts with the surrounding magnetic field, some helical structures show up (see the white box in Animation 1 around 09:30 UT) with bright plasmas swirling along the magnetic field lines.

Note that, despite two ribbon-like brightenings generated by the falling plasma, no notable flare (i.e., traditional {\em GOES} X-ray event) was observed during this failed eruption. The {\em GOES} 1--8 \AA~soft X-ray flux (the black dotted curve in Figure \ref{fig-obs}(a)) is basically flat at the $\sim$B3 level during the main eruption. In addition, no evident flare loops (i.e., filled with the plasma from chromospheric evaporation) show up in the EUV images. The integrated EUV light curves (Figure \ref{fig-obs}(a)) do not show a clear cooling trend as flare loops usually behave \citep{ying14,qiuj16}, either.

\section{Analysis and Results}

\subsection{the brightenings near the solar surface: impacts of \\the filament material channelled along the coronal field lines}
\label{sec-imp}

\begin{figure*}
\centering
\includegraphics[width=11cm]{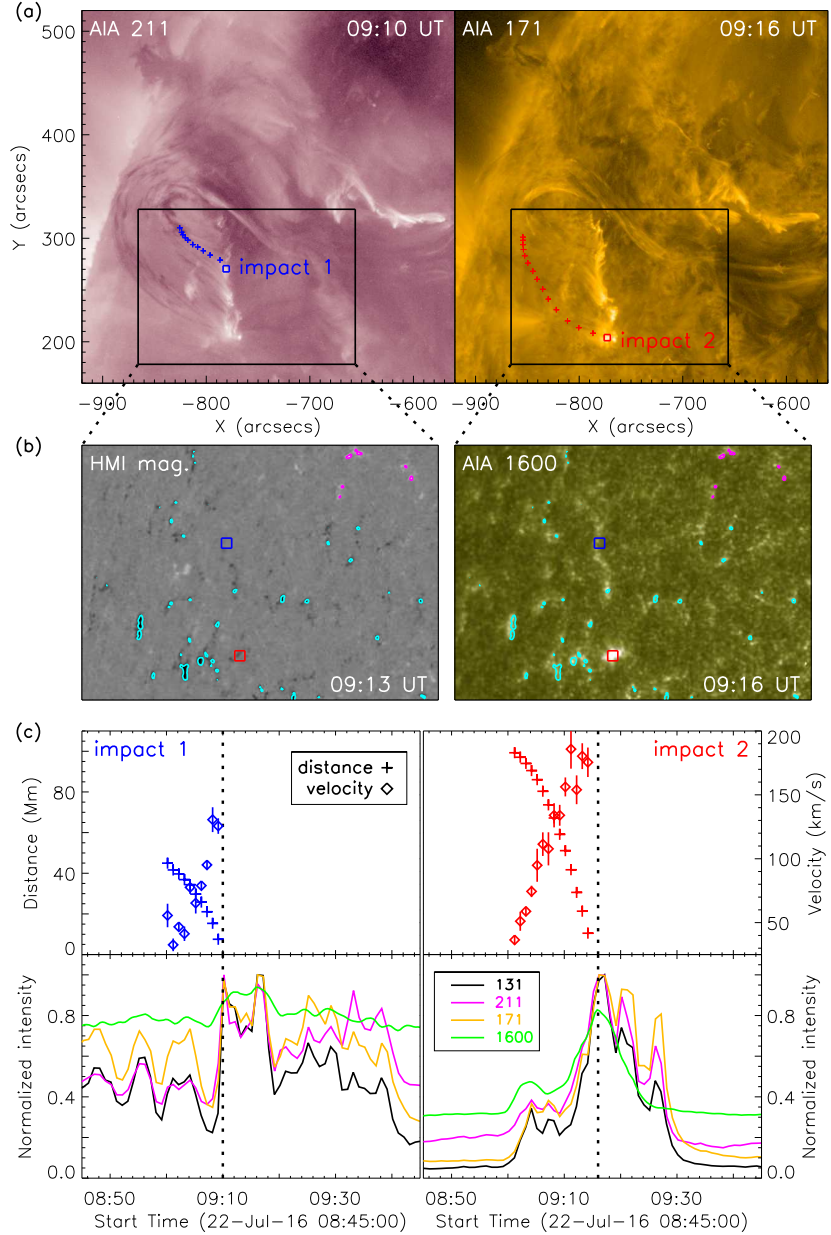}
\caption{{\small Two sample impacts (impact 1 and impact 2) of filament fragments. (a) AIA 211 \AA~and 171 \AA~images with the falling paths of the fragments tracked by plus symbols and their impact locations denoted by squares (with an area of 6\arcsec$\times$6\arcsec). The black box indicates the field of view of panel (b). (b) HMI magnetogram and AIA 1600 \AA~image with the two impact locations marked. The cyan and magenta contours represent the magnetic polarities at $-$100 and $+$100 G, respectively. (c) The upper panels show the falling paths (plus symbols; corresponding to the left scale that is defined as the distance to the impact location) and the velocities (diamonds; corresponding to the right scale) derived from the falling paths. The error in velocity represents the uncertainty from multiple AIA images. The lower panels plot the AIA light curves for the two impact brightenings (integrated over the squared area). The vertical dotted lines mark the impact times (09:10 UT for impact 1 and 09:16 UT for impact 2).}}
\label{fig-imp}
\end{figure*}

The multi-temperature AIA images show that the filament materials slide down along the lifting coronal field lines and hit the chromosphere causing two ribbon-like brightenings. Here we focus on these brightenings in the context of the material impacts first. From Figure \ref{fig-obs} and the accompanying Animation 1, one can find some obvious features in the impact brightenings. (1) The brightenings show up in all the EUV passbands as well as the UV 1600 \AA~channel almost simultaneously, though in the latter, the two ribbons are not well defined (see Figure \ref{fig-imp}(b)). This indicates that the impacting materials are rapidly heated to millions of degrees and have a wide temperature range. (2) The brightenings generally display patches or pieces in shape. (3) The brightenings fade out quickly with a typical lifetime of only several minutes. (4) The brightenings show multiple peaks in the EUV light curves at some locations (Figure \ref{fig-imp}(c)), implying that a series of impacts occur there.

For a further quantitative analysis, we select two sample locations (see the blue and red squares in Figure \ref{fig-imp}) on the ribbon-like brightenings, and track the falling paths of the impacting plasmas (denoted by plus symbols in Figure \ref{fig-imp} and the accompanying Animation 2\footnote{We derotate the AIA images when tracking the falling paths of the two sample impacts.}). From the falling paths, we derive the plane-of-sky velocities (diamonds in Figure \ref{fig-imp}(c)) that increase from tens to hundreds of km s$^{-1}$. As seen in Animation 2, the kinetic motion of the material actually contains two components, i.e., an expanding (or unshearing) motion and a falling one, as the filament fragments get elongated and channelled by the coronal field lines. When the dark fragments hit the chromosphere (at 09:10 UT for impact 1 and 09:16 UT for impact 2), they generate significant brightenings as shown in the EUV and UV light curves (Figure \ref{fig-imp}(c)) seemingly converting kinetic energy to thermal energy. We find that the impact velocities of the two sample locations (i.e., 126$\pm$6 and 176$\pm$12 km s$^{-1}$) are similar to the ones measured in \cite{gilb13}. Therefore, heating by kinetic energy dissipation of falling material is a viable mechanism to explain these observations.

The two ribbon-like brightenings support the gravitational model proposed for disappearing filaments by \cite{hyde67}. These brightenings correspond to the chromospheric footpoints of a row of coronal field lines, which is consistent with the distribution of the pair of magnetic polarities on the HMI magnetogram (see Figure \ref{fig-obs}(b) and Animation 1).

Note that, in this event, one can also find that some filament materials have brightened before they impact the chromosphere (such as the bright threads indicated by the black arrow in Figure \ref{fig-obs}). Moreover, some materials seem to mainly brighten in the corona. These materials are presumably heated and ionized by kinetic energy dissipation in the corona \citep{inne16} or by shock fronts and thermal conduction ahead of the falling materials \citep{petr16}.

\noindent \\ \\

\subsection{the brightenings in the helical structures: \\reconfiguration of the coronal magnetic field}
\label{sec-rec}

\begin{figure*}
\centering
\includegraphics[width=12cm]{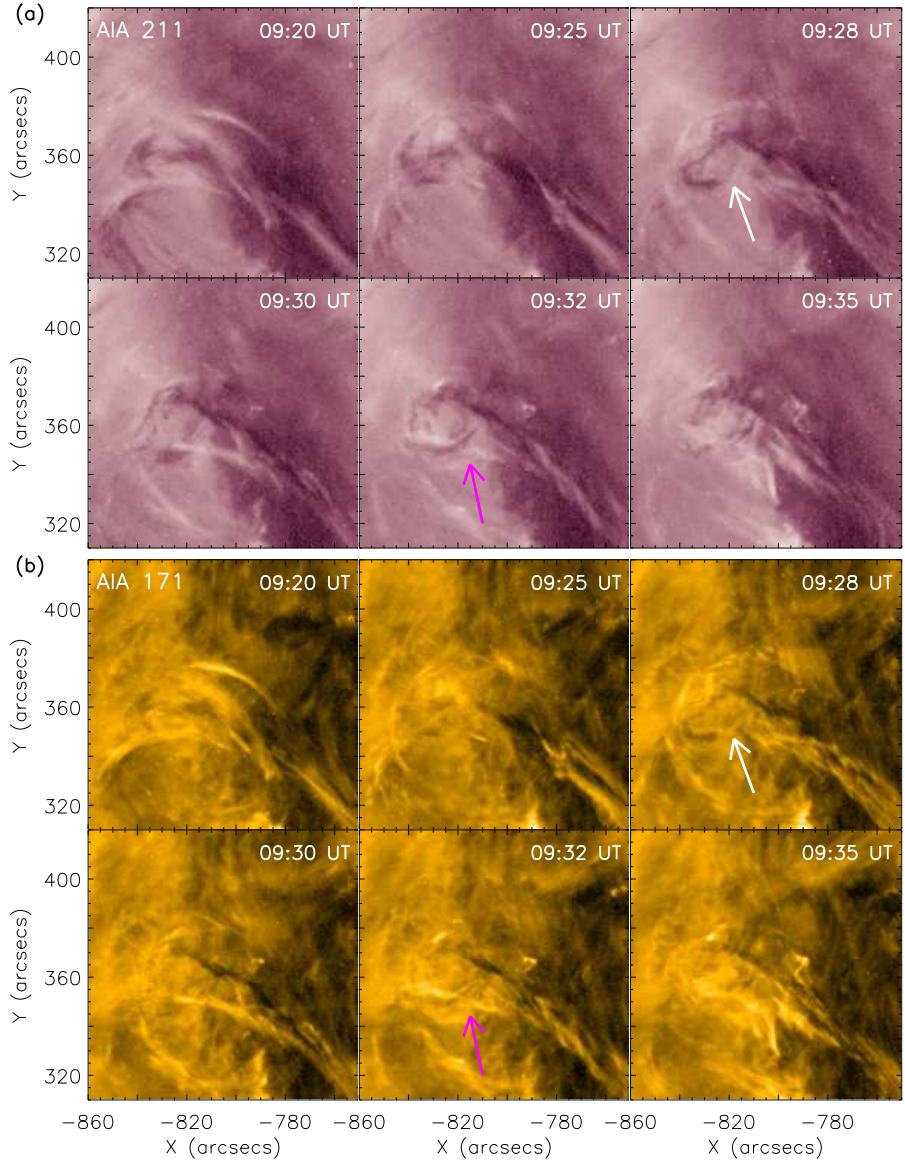}
\caption{{\small Temporal evolution of the helical structures in (a) AIA 211 \AA~and (b) 171 \AA~images. The white and magenta arrows indicate the helical structures.}}
\label{fig-hlc}
\end{figure*}

In this event, evident brightenings also show up in some helical (or twisted) structures which can be described as magnetic flux ropes, as shown in Animation 1 (see the white box). These brightenings are visible in almost all the EUV passbands, indicating that the plasma is heated to a few MK. The brightenings and the helical flux rope (indicated by the arrows in Figure \ref{fig-hlc}) probably suggest that magnetic reconnection occurs as the sheared filament arcade expands and interacts with the surrounding magnetic fields \citep{pare14,vial15}. This reconnection can heat the filament material and transform some sheared arcades into twisted field lines comprising the flux rope \citep{vanb89,long07,prie17}.

\begin{figure*}
\centering
\includegraphics[width=11.5cm]{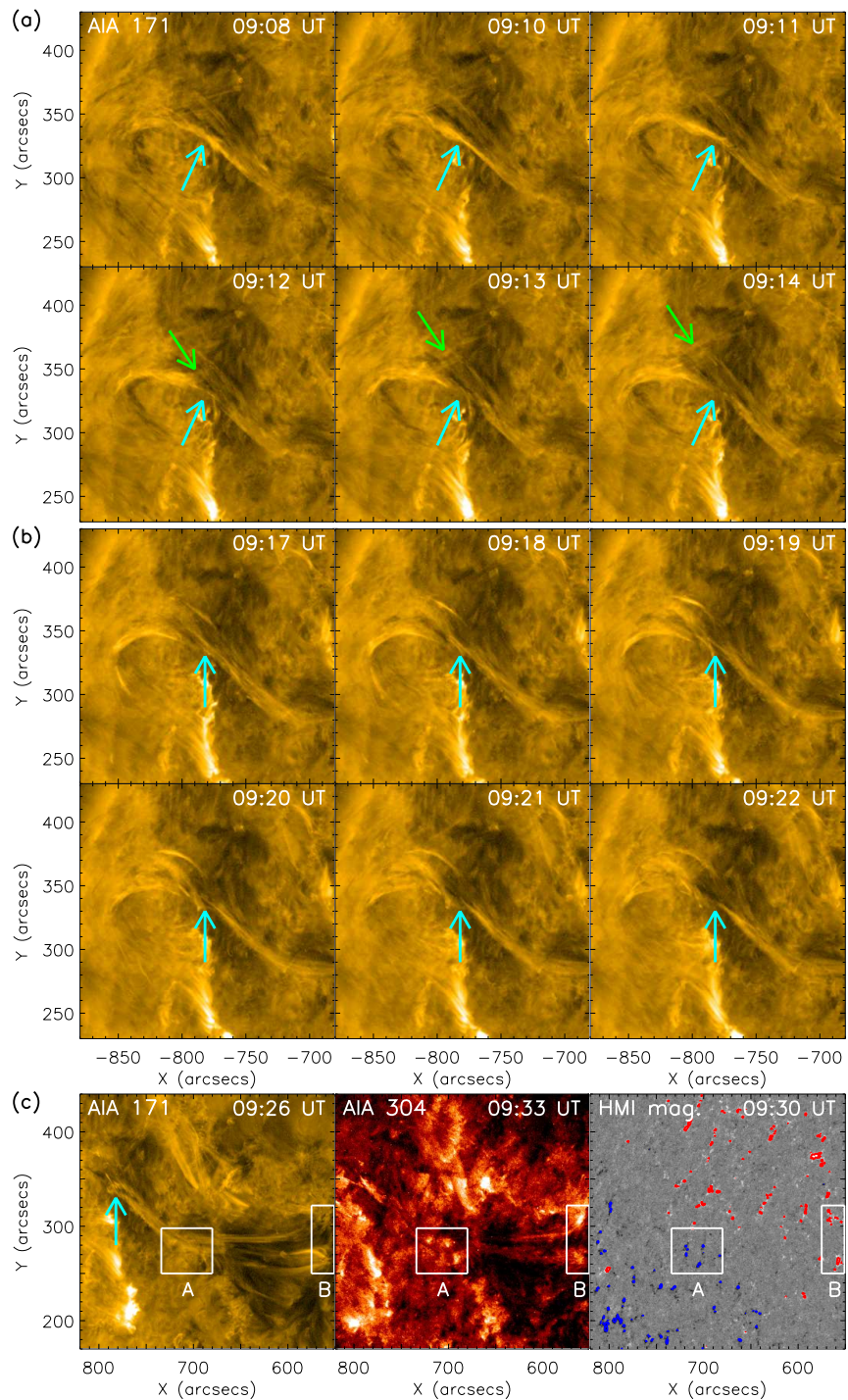}
\caption{{\small (a) and (b) Temporal evolution of the filament in AIA 171 \AA~showing the variation of plasma distribution. The cyan arrows indicate the break of the brightened filament plasmas. The green arrows in panel (a) mark the redirection of some brightened plasmas. (c) AIA 171 \AA~and 304 \AA~images and the HMI magnetogram showing a new connection of coronal field line footpoints. The white boxes A and B mark the footpoints of the reconnected field lines. The cyan arrow in the AIA 171 \AA~image denotes the same position as the cyan arrow in panel (b). The red and blue contours in the magnetogram represent the magnetic polarities at $+$100 and $-$100 G, respectively.}}
\label{fig-rec}
\end{figure*}

\begin{figure*}
\centering
\includegraphics[width=15cm]{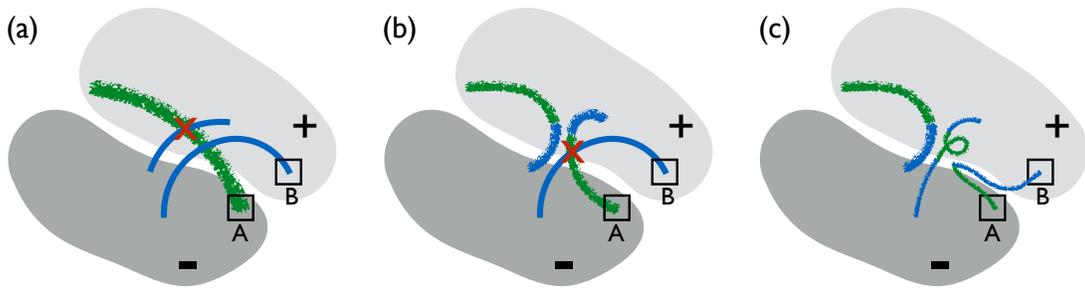}
\caption{{\small Sketch of the reconnection between the sheared filament arcade and the surrounding magnetic fields. (a) Pre-reconnection configuration. The green arc represents the sheared filament arcade. Two blue lines are surrounding magnetic field lines. (b) and (c) Post-reconnection configuration. The red crosses denote the reconnection sites. The plus and minus signs represent the positive and negative magnetic polarities, respectively. The boxes A and B mark the footpoints of reconnected field lines, as described in the context for the observed new connection of footpoints. Note that here we only show the field lines being involved in the magnetic reconnection. It seems that in observations only part of the filament arcade is involved in the reconnection.}}
\label{fig-ske}
\end{figure*}

More evidence of magnetic reconnection comes from a break of filament plasma distribution (basically tracing the coronal magnetic field) and a new connection of field line footpoints. From Figures \ref{fig-rec}(a) and (b) and the accompanying Animation 3, one can see that the brightened filament plasmas seem to break up (indicated by the cyan arrows) as time evolves. In particular, during the break at an early time, some of the brightened plasmas are also redirected (marked by the green arrows in Figure \ref{fig-rec}(a); also see Animation 3). Near the break site, some plasmas begin to swirl up along the helical structures and some others fall down along the coronal (i.e., reconnected) field lines. The break of filament plasma distribution then brings a new connection of the footpoints of coronal field lines, as shown in Figure \ref{fig-rec}(c). It is seen that one of the footpoint (in the white box A; also denoted in Figure \ref{fig-obs}(b)) of the sheared filament arcade is connected to a new footpoint (in the white box B), both of which start to brighten at the same time. The footpoints in A and B, as clearly seen in the AIA 304 \AA~images, are located at the negative and positive magnetic polarities on the HMI magnetogram, respectively. The magnetic configuration of reconnection between the sheared filament arcade and the surrounding magnetic fields is sketched in Figure \ref{fig-ske}.

\subsection{the brightenings on top of the erupting magnetic fields: \\rarefaction of the plasma constrained in the coronal magnetic field}
\label{sec-rar}

\begin{figure*}
\centering
\includegraphics[width=16cm]{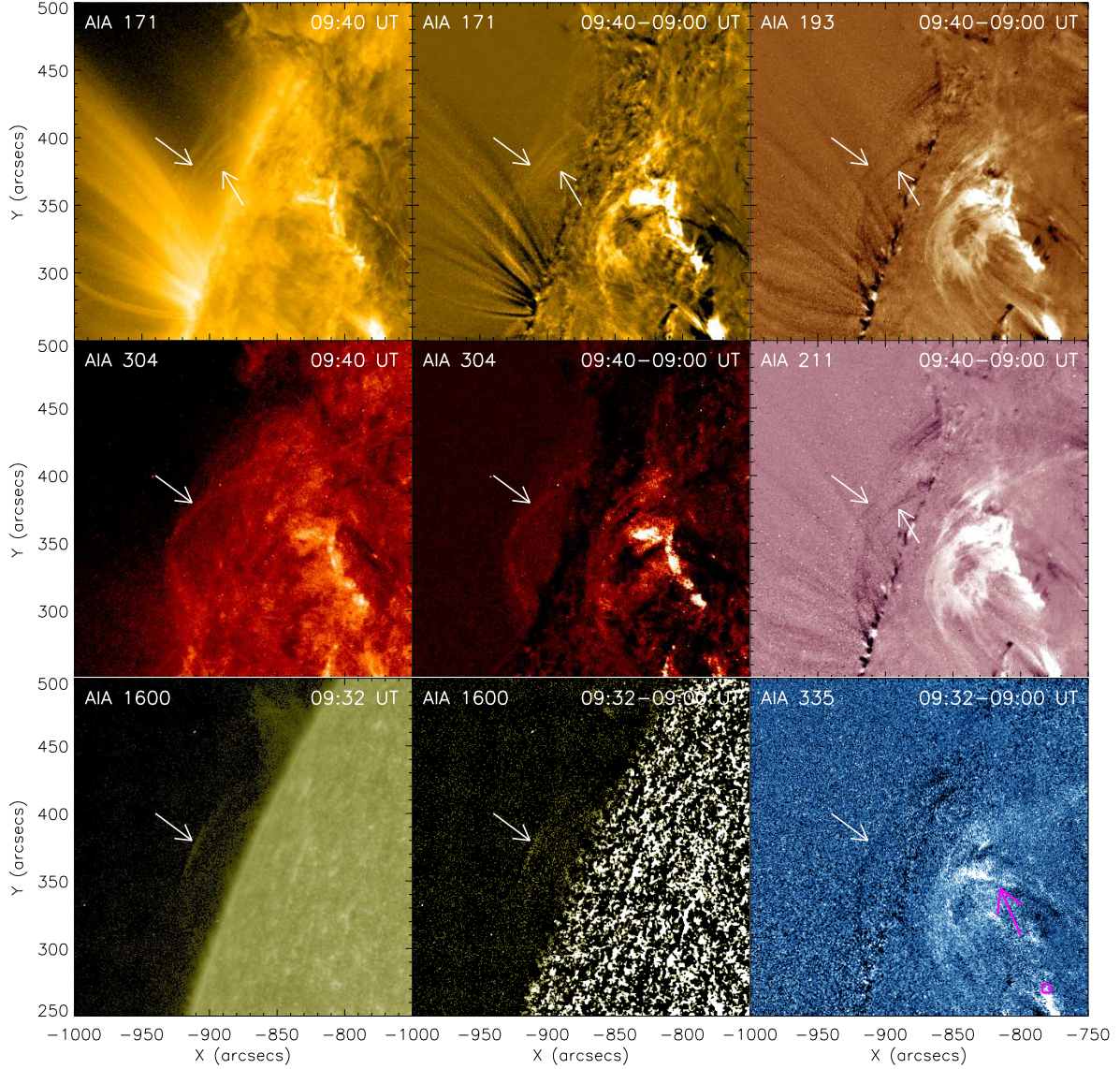}
\caption{{\small Left column: AIA images in three low-temperature passbands ($<$1 MK). Middle column: difference images of the left-column ones. Right column: difference images in three high-temperature passbands ($>$1 MK). The white arrows indicate the brightened threads that display either bright or dark features in different passbands. The magenta arrow and square in the AIA 335 \AA~image are the same as the ones marked in Figures \ref{fig-hlc} and \ref{fig-imp}(a) for the helical structure and the location of impact 1, respectively.}}
\label{fig-rar}
\end{figure*}

Besides the brightenings at the two ribbon-like impacts and in the helical structures, we observe thread-like brightenings on top of the erupting magnetic fields (see the white arrows in Figure \ref{fig-rar} and the accompanying Animation 4). These weak emissions only show up as bright threads in some of the lower-temperature passbands ($<$1 MK), i.e., in the AIA 1600 \AA, 304 \AA, and 171 \AA~passbands. While in the relatively hot 193 \AA, 211 \AA, and 335 \AA~passbands ($>$1 MK), these threads display dark features, indicative of a relatively low temperature there. This is quite different from the brightenings due to material impacts (S\ref{sec-imp}) and magnetic reconnection (S\ref{sec-rec}), which are visible in almost all the EUV passbands covering a temperature range up to a few MK (see the magenta arrow and square overplotted in Figure \ref{fig-rar}). This implies that a different mechanism is at work in the thread-like brightenings on top of the erupting magnetic fields.

From Animation 4, we see that the thread structures filled with dark filament materials appear to expand and brighten up in the low-temperature passbands. The expansion of the thread structures and, in particular, the falling down of plasma along the field lines (described in S\ref{sec-imp}), can make the plasma near the top rarefied. We therefore propose a possible cause of the thread-like brightenings in terms of an energy imbalance between the radiative cooling and the background heating. Suppose that originally the radiative cooling is balanced by the background heating from, say, wave damping \citep{erdl04,anto08} or nanoflare heating \citep{park88,klim06}, on top of the erupting magnetic fields. If the mass density there decreases, the radiative cooling rate, proportional to the density square, drops more rapidly than the background heating rate, which is usually proportional to the density. Then, an energy imbalance could induce a heating effect there. This process has been reproduced in numerical simulations (P. F. Chen 2016, private communications). Since such a heating only works subtly, the induced temperature increase should be small and probably only visible in some low-temperature passbands, as revealed in this event.

\section{Summary and Discussions}

We have presented a failed filament eruption in a quiet-Sun region observed by AIA and HMI. In this event, the erupting filament is confined by the surrounding magnetic field. When interacting with the ambient magnetic field, the filament material brightens up and flows along the magnetic field lines through the corona to the chromosphere. We find two ribbon-like brightenings near the solar surface, plasma heating in some helical structures, as well as thread-like brightenings on top of the erupting magnetic fields. Based on their different observational features, these brightenings in different magnetic structures likely result from different kinds of heating mechanisms.

First, the two ribbon-like brightenings near the solar surface mainly come from the impacts of filament fragments through kinetic energy dissipation \citep{inne16,petr16}. The fragments slide down along the lifting coronal field lines and crash into the chromosphere to cause impact brightenings. Although the impact is most likely the cause here, we do not exclude some other mechanisms, such as a reconnection between the magnetic field channelling the impact material and the low-lying loops \citep{gilb13}, which might play some role in these brightenings. 

Second, the brightenings in the helical structures are likely from the magnetic reconnection between the sheared filament arcade and the surrounding magnetic field. Although we find some probable signatures of reconnection, such as brightenings, helical structures, and breaks of plasma distribution as well as a new connection of coronal field line footpoints, it is interesting that no evident flare loops were observed in this event. Moreover, magnetic reconnection usually accelerates non-thermal particles to produce hard X-ray emissions; however, no excess emissions were found in the hard X-ray light curves from {\em RHESSI} during this event. The above facts could be explained by the following reasons. (1) The reconnection between the sheared filament arcade and the surrounding magnetic field probably involves only a little magnetic flux (so as not to produce any notable flare) in this quiet-Sun region with a weak magnetic field, and thus could not produce evident chromospheric evaporation to fill and brighten the flare loops. In fact, we notice that \cite{vand14} reported a similar event in which a reconnection occurs between the erupting flux rope and the ambient magnetic field. Their event shows little signature of chromospheric evaporation in flare loops, just like our case. (2) The energy released from the reconnection might be mainly converted to plasma heating and only a small fraction is used to accelerate electrons. Those non-thermal electrons, if present, may be quickly thermalized by the heated dense filament plasma; thus, no detectable hard X-ray emission could be produced. This case has been reported by \cite{mill08} in which a microflare showed hot plasma with absence of any detectable hard X-ray emission. It is also noteworthy that, in addition to the energy released through reconnection, compression of the filament plasma may also provide some heating energy. 

Finally, for the bright threads on top of the erupting magnetic fields only appearing in the low-temperature passbands, a possible explanation is the energy imbalance from a fast drop of radiative cooling due to the plasma rarefaction. Such an effect has rarely been reported in previous studies. This failed filament eruption is a promising example showing this mechanism to work at some places. We do not, however, exclude other possibilities, such as plasma compression by the expanding thread structures, which could be responsible for these thread-like brightenings. 

Overall, this event reveals different kinds of brightenings resulting from different heating mechanisms. In the future, a numerical simulation is needed to show how these mechanisms work jointly in a single event.


\acknowledgments
{\em SDO} is a mission of NASA's Living With a Star Program. The authors are grateful to Pengfei Chen and Jiong Qiu for valuable discussions and thank the referee for constructive comments that significantly improve the manuscript. Y.L. and M.D.D. are supported by NSFC under grants 11373023 and 11403011, and by NKBRSF under grant 2014CB744203. Y.L. is also supported by the Postdoctoral Science Foundations from Jiangsu Province and China Postdoctoral Office, by the Fundamental Research Funds for the Central Universities, and by the Office of China Postdoctoral Council (No. 29) under the International Postdoctoral Exchange Fellowship Program 2014. The work at MSU is supported by NSF grant 1460059.

\bibliographystyle{apj}

\end{document}